\documentclass[showpacs,showkeys,preprintnumbers]{revtex4}%
\usepackage{amsfonts}
\usepackage{epsfig}
\usepackage{amsmath}
\usepackage{graphicx}
\usepackage{dcolumn}
\usepackage{bm}
\usepackage{amssymb}%
\providecommand{\U}[1]{\protect\rule{.1in}{.1in}}

\begin{document}
\title{Wheeler-DeWitt Universe Wave Function in the presence of stiff matter}
\author{Francesco Giacosa$^{\text{(a,b)}}$}
\email{fgiacosa@ujk.edu.pl}
\author{Giuseppe Pagliara$^{\text{(c)}}$}
\email{pagliara@fe.infn.it}
\affiliation{$^{\text{(a)}}$Institute of Physics, Jan Kochanowski University, ul.
Swietokrzyska 15, 25-406 Kielce, Poland,}
\affiliation{$^{\text{(b)}}$Institute for Theoretical Physics, Goethe University,
Max-von-Laue-Str.\ 1, 60438 Frankfurt am Main, Germany.}
\affiliation{$^{\text{(c)}}$Dip. di Fisica e Scienze della Terra dell'Universit`a di
Ferrara and INFN Sez. di Ferrara, Via Saragat 1, I-44100 Ferrara, Italy}

\begin{abstract}
We study the Wheeler-DeWitt (WDW) equation close to the Big-Bang. We argue
that an interaction dominated fluid (speed of sound equal to the speed of
light), if present, would dominate during such an early phase. Such a fluid
with $p=\rho\propto1/a^{6}$ generates a term in the potential of the wave
function of the WDW equation proportional to $-1/a^{2}.\ $This very peculiar
quantum potential, which embodies a spontaneous breaking of dilatation
invariance, has some very remarkable consequences for the wave function of the
Universe: $\Psi(a)$ vanishes at the Big-Bang: $\Psi(0)=0$; the wave function
$\Psi(a)$ is always real; a superselection rule assures that the system is
confined to $a\geq0$ without the need of imposing any additional artificial
barrier for unphysical negative $a$. These results do not depend on the
operator-ordering problem of the WDW equation.

\end{abstract}

\maketitle

\textit{Motivation:} At the very beginning of the Universe evolution, just
after the big-bang, the energy density was extremely high. In a classical
treatment, one has the so-called big-bang singularity: the energy density
diverges when the scale factor $a$ defining the Friedmann-Robertson-Walker
(FRW) metric vanishes. It is however well known that `classical' general
relativity (GR) is not sufficient, since at such an early stage the influence
of Quantum Mechanics (QM) is expected to be large.

One of the first treatments of quantum gravity was put forward by\ Wheeler and
DeWitt \cite{wdw}: it is a canonical approach, in which the Hamiltonian of
general relativity is quantized, hence the wave function is a function of the
(spacial) metric. A Schr\"{o}dinger-like equation, called Wheeler-DeWitt (WDW)
equation, emerges. Although we still do not know if this is the correct and/or
the most efficient way to quantize gravity
\cite{kiefer,kiefersandhoefer,rovelli}, it represents a useful approach to
describe various problems in which both GR and QM merge.\ This is especially
the case of quantum cosmology.

The WDW equation simplifies tremendously when a uniform and homogeneous FRW
Universe is considered: the wave function is solely function of the scale
parameter $a$ [hence, $\Psi=\Psi(a)$], see e.g. \cite{kolbturner,atkatz}.
However, it is not clear what fixes the boundary conditions (if any)
associated with the WDW equation and a long debate has emerged in this
context: while Hartle and Hawking find, within the so-called no-boundary
proposal, a real wave function \cite{hh} (see also
\cite{hawking2,hertog,dorronsoro}) containing both ingoing and outgoing waves,
Vilenkin \cite{vilenkin} put forward a complex wave function corresponding
solely to an outgoing wave. Usually, in such studies of the early time of the
Universe, only the curvature and the constant cosmological terms are retained.
For a recent description of the other possible components, such as matter and
radiation, see Refs. \cite{vieira,belinchon}. Indeed, the interest on the wave
function of the Universe is very strong, as the recent lively and vibrating
dispute on the effect of quantum gravitational perturbations in the early
universe shows \cite{feldbrugge,dorronsoro2,feldbrugge2}.

Besides the problem of the explicit form of the wave function mentioned above,
there are other issue connected to the WDW equation: (i) What should be the
wave function at the big-bang, $\Psi(0)$? It is non-vanishing for both the
Hartle-Hawing and Vilenkin solutions mentioned above. (ii) How to implement
the classical constraint $a\geq0$ \cite{isham}? Usually, the transformation
$a=e^{\Omega}$ is performed \cite{kiefer}, but this is merely a mathematical
trick. (iii) Should the wave function be real or complex? (iv) Is there any
influence of the so-called operator ordering problem?

In this work, we shall consider the effect of a stiff-matter interaction
dominated gas, for which the pressure equalize the energy density, $p=\rho$
(the speed of sound is $c,$ hence maximal).

Namely, whatever d.o.f. are present in the very early universe, their strong
interactions could generate such a gas. Quite remarkably, recent studies on
the most dense form of strongly interacting matter, taking place in the core
of neutron stars, show that the speed of sound should be larger than
$\sqrt{1/3}$ (the value for a non-interacting ultrarelativistic gas of
massless particles) in order to explain the existence of massive neutron stars
\cite{bedaque}. The study of this causal limit is common in neutron stars,
since it is the stiffest possible Equation of State (EoS) and it is useful to
set the limit to the highest possible mass
\cite{Rhoades:1974fn,Koranda:1996jm,neutronstarnc}. Theoretically, it is well
understood how such a gas emerges due to strong interactions
\cite{glendenning}. Indeed, a component of the form $p=\rho$ has been taken
into account for classical cosmologies (e.g. Refs. \cite{nemiroff,roulettes}),
but not in their quantum version. Moreover, an equivalent term in the
classical FRW equation appears also in the presence of so-called kination
domain \cite{palliskination,kination}, in which a massless (and homogenous)
scalar field $\phi(t)$ is considered. Yet, the quantized version is pretty
different (since the wave function depends on $a$ and $\phi$ as well) and one
should consider that a massless scalar field $\phi(t,\mathbf{x})$ should
actually generate a gas of the type $p=\rho/3$ when fluctuations are included.

In the realm of quantum cosmology, a fluid with $p=\rho$ corresponds to a term
of the type $-1/a^{2}$ in the effective potential of the WDW equation. Such a
potential, if present, necessarily dominates at small $a$ (other possibilities
are excluded since would violate causality). This is indeed a \textit{very
peculiar quantum} potential that breaks all our naive expectations for a
quantum system, see Ref. \cite{griffiths} and also refs.
\cite{comment,others1overxsquared}. At a first sight, it seems that no bound
state should exist, since, if one exists, a continuum of bound states, one for
each negative energy, would also exist. At a closer inspection, the system is
much more interesting and its detailed treatment imposes to render the
Hamiltonian self-adjoint \cite{griffiths,selfadjoint}: if the attraction is
below a certain critical value, there is a single bound state, but, above,
there is an infinity of bound states (one of which with lower energy). In
turn, this system provides a beautiful example of an anomaly: a characteristic
length in the system emerges, which is in a sense similar to the development
of the Yang-Mills energy scale in QCD.\textbf{ }

Quite remarkably, the unexpected features of the potential $-1/a^{2}$ in the
WDW equation may help to relieve in an elegant way the problems of the WDW
approach listed above: (i) The wave function vanishes at the big-bang:
$\Psi(a=0)=0.$ This condition reminds the old idea of DeWitt according to
which a vanishing wave function could represent a solution of the problem of
the singularity \cite{wdw,kiefer}. (ii) It generates a superselection rule
according to which only positive (or negative) values of $a$ are allowed.
Hence, once $a>0$ is chosen, the wave function is automatically nonzero only
on the r.h.s. and there is no need of any further artificial restriction.
(iii) The wave function is \textit{real} in agreement with the result of
Hartle and Hawking \cite{hh}.\ (iv) The results are qualitatively independent
on (a very large class of) choice(s) of the operator ordering.

\bigskip

\textit{WDW equation in cosmology:} We briefly review how the WDW equation
emerges in cosmology. First, we consider the scale factor $a\equiv a(t)$ as a
field with dimension length subject to the classical Lagrangian%
\begin{equation}
L_{\text{FRW}}=-Ca^{3}\left[  \left(  \frac{\dot{a}}{a}\right)  ^{2}%
-\frac{kc^{2}}{a^{2}}+\frac{\Lambda c^{2}}{3}+\frac{8\pi G}{3c^{2}}%
\rho\right]  \text{ with }C=\frac{3\pi c^{2}}{4G},
\end{equation}
where $k$ and $\Lambda$ parametrize the curvature and the cosmological
constant contributions to the Universe's evolution.

The energy density $\rho$ describes the contribution of matter and energy.
Here, we shall consider that each component fulfills the EoS $p=w\rho,$ which
has a constant speed of sound $v_{sound}=c\sqrt{dp/d\rho}=c\sqrt{w}\leq c$.
The adiabatic expansion $dE+pdV=0$ translates into $d(\rho a^{3})+pd(a^{3}%
)=0$, then:
\begin{equation}
a\frac{d\rho}{da}=-3(\rho+p)=-3(1+w)\rho\implies\rho(a)=\frac{A_{w}%
}{a^{3(w+1)}}\text{ .}%
\end{equation}
As renowned \cite{cosmologyreview}, for $w=0$ a Universe dominated by dust is
obtained ($\rho_{\text{dust}}\propto a^{-3},$ dark plus visible matter, about
$30\%$ of contribution to the present state of the expansion, the rest being
the present cosmic inflation). A radiation dominated Universe is found by
setting $w=1/3$ ($\rho_{\text{radiation}}\propto a^{-4}$); this was relevant
in the radiation dominated era of the Universe. Of course, the use of a
constant $w$ is an approximation, since a relativistic plasma with
$w\simeq1/3$ turns into a nonrelativistic gas $w\simeq0$ when the Universe
cools down. Moreover, at a given time, different disjunct components of the
fluid can follow their own EoS, leading to
\begin{equation}
\rho=\rho_{\text{dust}}+\rho_{\text{radiation}}+...
\end{equation}

Here, we argue that at the very beginning of the Universe, an interaction
dominated gas whose EoS is given by $w=1$ could have been present (whatever
d.o.f. were relevant, see e.g. Ref. \cite{neubert} and refs. therein). For
this fluid:
\begin{equation}
p=\rho\rightarrow v_{sound}=1\text{ and }\rho_{\text{int-dom}}=\frac
{A_{\text{int-dom}}}{a^{6}}.
\end{equation}
Clearly, this component can be relevant only at a very stage of the expansion,
since (i) it decreases very fast for increasing $a$ and (ii) the strong
interaction generating it weakens down and transforms this fluid into a more
conventional component [hence, $w$ decreases from $1$ to $1/3$ (or even smaller)].

The first Friedmann equation is obtained by imposing that the Hamiltonian
vanishes:
\begin{equation}
H_{\text{FRW}}=p\dot{a}-L_{\text{FRW}}=0\text{ with }p=\frac{\partial
L_{\text{FRW}}}{\partial\dot{a}}-2Ca\dot{a}\text{ .} \label{constr}%
\end{equation}
This constraint follows from the invariance under coordinate transformations
of GR. In terms of $a$ and $\dot{a},$ Eq. (\ref{constr}) gives the first
Friedmann equation%
\begin{equation}
\left(  \frac{\dot{a}}{a}\right)  ^{2}+\frac{kc^{2}}{a^{2}}-\frac{\Lambda
c^{2}}{3}-\frac{8\pi G}{3c^{2}}\rho=0\text{ .}%
\end{equation}
(The second Friedmann equation is obtained by studying the equation of motion
of $L_{\text{FRW}}$ together with the continuity equation, see details in
\cite{kolbturner,belinchon,vieira}). As function of $p$ and $a$, the
Hamiltonian reads:
\begin{equation}
H_{\text{FRW}}=\frac{-1}{4C}\frac{p^{2}}{a}+C\left(  -kc^{2}a+\frac{\Lambda
c^{2}}{3}a^{3}+\frac{8\pi G}{3c^{2}}\rho a^{3}\right)
\end{equation}
When promoting $H_{\text{FRW}}$ as an operator via $a\rightarrow a$ and
$p\rightarrow-i\hslash\partial_{a}$ and by choosing the ordering $\frac{p^{2}%
}{a}=\frac{1}{a}p^{2}$ (the following results do not depend on this choice as
we will explain later) one obtains the stationary Schr\"{o}dinger equation
with zero energy:
\begin{equation}
\left[  -\hslash^{2}\frac{d^{2}}{da^{2}}+V_{eff}(a)\right]  \Psi(a)=0\text{
with }V_{eff}(a)=4C^{2}\left(  kc^{2}a^{2}-\frac{\Lambda c^{2}}{3}a^{4}%
-\frac{8\pi G}{3c^{2}}\rho a^{4}\right)  \text{ .}%
\end{equation}
This is the famous WDW equation. It is a timeless equation: a discussion about
the emergence of time can be found in the literature
\cite{rovelli,isham,anderson,pintoneto}.

Here we are interested in the very early time evolution, therefore we consider
$\rho=\rho_{\text{int-dom}}=A_{\text{int-dom}}/a^{6}$ (we neglect dust and
radiation as well as other contributions, which become important at later
stages of the evolution). Thus, our final form for the effective potential
reads:%
\begin{equation}
\text{ }V_{eff}(a)=4C^{2}\left(  kc^{2}a^{2}-\frac{\Lambda c^{2}}{3}%
a^{4}\right)  -\frac{\alpha\hslash^{2}}{a^{2}}\text{ .}%
\end{equation}
The first term in the parenthesis is the one usually studied for the early
quantum cosmology \cite{hh,hawking2,vilenkin,gibbons}, and the second piece
represents the additional part being the main subject of the present work. It
is parametrized by the dimensionless coupling $\alpha$%

\begin{equation}
\alpha=4C^{2}\frac{8\pi G}{3c^{2}\hslash^{2}}A_{_{\text{int-dom}}}=6\pi
^{3}\frac{c^{2}}{G\hslash^{2}}A_{_{\text{int-dom}}}\text{.} \label{alfa}%
\end{equation}
Thus, for $a$ very small, the term $-\alpha\hslash^{2}/a^{2}$ dominates,
leading to the WDW equation
\begin{equation}
\left[  \frac{d^{2}}{da^{2}}+\frac{\alpha}{a^{2}}\right]  \Psi(a)=0\text{ (for
}a\text{ very small) .} \label{WDWsmalla}%
\end{equation}
The potential $-1/a^{2}$ has very remarkable properties that have been studied
in detail in Refs. \cite{griffiths,comment,others1overxsquared}. Since
$\alpha$ in Eq. (\ref{alfa}) is dimensionless, there is no typical energy
scale in the problem: writing the eigenvalue equation for this potential, one
can infer that --if it admits a bound state-- there are infinite bound states,
or in other terms there is no ground state. By indicating with $k_{e}^{2}$ the
eigenvalue of the operator on the l.h.s. of Eq. (\ref{WDWsmalla}) (which, in
our case is set zero in order to recover the WDW equation of (\ref{WDWsmalla}%
)), one can show that $\Psi(a)\propto\sqrt{a}K_{ig}(k_{e}a)$, where
$g=\sqrt{\alpha-1/4}$ and $K_{ig}$ is the modified Bessel function of order
$ig$ \cite{griffiths}.
Then, one concludes that the wave function $\Psi(a)$ vanishes in $a=0$:
\begin{equation}
\Psi(a=0)=0\text{ .}%
\end{equation}
This is a first important and general result of our study: when considering an
interaction dominated fluid that could have appeared just after the big bang,
the wave function of the Universe fulfills the requirement postulated long ago
by DeWitt to solve the problem of the big-bang singularity. A second
properties of the $1/a^{2}$ attractive potential concerns a superselection
rule imposed on the allowed range of the variable $a$. As shown in
\cite{comment} the quantum system is confined to $a>0$ (or $a<0$)
\cite{comment}; in other words, there is no linear superposition of wave
functions which live at $a>0$ with the ones at $a<0$. Hence, there is no
problem with negative values of $a$.

Next, we turn to explicit solutions in order to show that the wave function is
real. When interpreting $H_{\mathrm{FRW}}$ as an operator associated to a
physical observable (i.e. the Hamiltonian) one has to verify that the operator
is a self-adjoint operator i.e. it is symmetric and the domain of it coincides
with the domain of its adjoint. As discussed in \cite{griffiths}, the property
of self-adjointness for the $-1/a^{2}$ potential is obtained by imposing a
specific boundary condition on the wave function (see Eq. (73) and (81) of
Ref. \cite{griffiths}, see also Ref. \cite{selfadjoint}). For $g=\sqrt
{\alpha-1/4}\neq0$,
\begin{equation}
\sqrt{a}\left[  e^{2ig\log\frac{a}{a_{0}}}-1\right]  \frac{d\Psi^{\ast}%
(a)}{da}-\frac{1}{\sqrt{a}}\left[  \left(  \frac{1}{2}+ig\right)
e^{2ig\log\frac{a}{a_{0}}}-\left(  \frac{1}{2}+ig\right)  \right]  \Psi^{\ast
}(a)\rightarrow0\text{ ,} \label{sa1}%
\end{equation}
while for $g=\sqrt{\alpha-1/4}=0$%
\begin{equation}
\sqrt{a}\log\frac{a}{a_{0}}\frac{d\Psi(a)}{da}-\frac{1}{\sqrt{a}}\left[
1+\frac{1}{2}\log\frac{a}{a_{0}}\right]  \Psi(a)\rightarrow0\text{ .}
\label{sa2}%
\end{equation}

The general solution of Eq. (\ref{WDWsmalla}) is a linear combination of two
independent functions with, in general, complex coefficient. However, when
imposing Eqs. (\ref{sa1}) and (\ref{sa2}) for the cases $g\neq0$ and $g=0$
respectively, one finds that only real wave functions are allowed. Let us
discuss the case $g=0$: the solution is $\Psi(a)=\sqrt{a}(c_{1}+c_{2}%
\log(a/a_{0}))$ where $c_{1}$, $c_{2}$ are complex numbers. Eq. (\ref{sa2})
leads to the conditions $c_{1}=0$ and $c_{2}$ can be chosen as a real number,
thus the wave function is real. Similarly, one can compute the solution in the
general case $g\neq0$: $\Psi(a)=\sqrt{a}e^{-ig}(c_{1}+c_{2}x^{2ig})$ and
impose the limit of Eq. (\ref{sa1}). After a straightforward calculation, the
solutions for the wave function very close to the big-bang can be recasted in
the following compact form:%
\begin{equation}
\Psi(a)=\left\{
\begin{array}
[c]{c}%
N\sqrt{a}\sin\left[  g\ln(a/a_{0})\right]  \text{ }\\
N\sqrt{a}\ln(a/a_{0})\text{ }\\
N\sqrt{a}\left[  \left(  a/a_{0}\right)  ^{-\tilde{g}}-\left(  a/a_{0}\right)
^{\tilde{g}}\right]  \text{ }%
\end{array}%
\begin{array}
[c]{c}%
\text{for }g>0\text{ (i.e., }\alpha>1/4)\\
\text{for }g=0\text{ (i.e., }\alpha=1/4)\\
\text{for }g=i\tilde{g}>0\text{ (}0<\tilde{g}<1/2,\text{ i.e. }0<\alpha
<1/4\text{)}%
\end{array}
\right.  \label{psisol}%
\end{equation}
These solutions summarize the results of this paper. As anticipated
previously, $\Psi(a=0)=0$: the wave function vanishes at the big-bang, thus
offering a possible solution of the singularity problem (this is why $\alpha$
cannot be negative, otherwise $\Psi(a),$ although formally still given by the
last line of Eq. (\ref{psisol}), would be divergent for $a\rightarrow0$).
Moreover, we also have that $\Psi(a=a_{0})=0$, i.e., the wave function also
vanishing at the length emerging via the process of anomalous breaking of
dilatation symmetry. However, the value of the constant $a_{0}$ cannot be
determined. At first, it seems natural to set $a_{0}\simeq\sqrt{\frac{\hslash
G}{c^{3}}}=l_{P}\simeq10^{-33}$ cm, but there is actually no compelling reason
for that. The eventual role of this new fundamental length $a_{0}$ should be
investigated in the future. The constant $N$ is a normalization constant which
can be always taken as real, therefore $\Psi(a)$ is real. There is no problem
for $a\rightarrow0$, and also no problem for $a<0$ (it never goes to $a<0$
without imposing any additional requirement).

Finally, it is important to show that our main outcomes do not depend on the
ordering of the operators. Other prescriptions induce a shift of the critical
value of $\alpha$, but there is no qualitative change of our discussion. For
instance, for the choice $\frac{p^{2}}{a}\rightarrow\hat{p}\frac{1}{a}\hat{p}$
(as in \cite{vilenkin}) one still finds that $\Psi(a\rightarrow0)=0$ for each
$\alpha>0.$ Upon defining $\Psi^{new}(a)=\Psi(a)/\sqrt{a}$, Eq.
(\ref{WDWsmalla}) is re-obtained for a shifted $\alpha$:%
\begin{equation}
\left[  \frac{d^{2}}{da^{2}}+\frac{\alpha^{new}}{a^{2}}\right]  \Psi
^{new}(a)=0\text{ .} \label{psinew}%
\end{equation}
with $\alpha^{new}=\alpha-\frac{3}{4}$. Thus, being $\Psi^{new}(a)$ real,
$\Psi(a)$ is also such. For the critical value $\alpha^{new}=1/4$ (hence,
$\alpha=1$), $\Psi(a)=Na\ln(a/a_{0})$. (For $\alpha<3/4$, $\alpha^{new}<0$,
$\Psi^{new}(a)$ is still given as the last line of Eq. (\ref{psisol}), but for
$\tilde{g}>1/2,$ hence $\Psi^{new}(a\rightarrow0)$ diverges. There is however
no problem since this divergence is compensated by $\sqrt{a}$ as long as
$\alpha>0.$ The wave function $\Psi(a)$ always vanishes at the big-bang for
positive $\alpha$).

More in general, let us consider the two-parameter parametrization
$\frac{p^{2}}{a}\rightarrow\frac{1}{a^{i}}\hat{p}\frac{1}{a^{j}}\hat{p}%
\frac{1}{a^{i+j-1}}$ \cite{steigl} (our case corresponds to $i=1,$ $j=0,$
Vilenkin's choice to $i=0,$ $j=1,$ the parametrization $j=1-i$ was studied in
Ref. \cite{kontoleon,kolbturner}). The very same steps can be performed. The
wave function is such that $\Psi(a\rightarrow0)=0$ for any $\alpha>0$ as long
as $i\geq0,j\geq0$ and $i+j\leq1$ (intuitively, we split $1/a$ into three
parts, each one with a positive power in the denominator). A redefinition
$\Psi^{new}(a)=a^{-\delta}\Psi(a)$ for which Eq. (\ref{psinew}) holds for
$\delta$ and $\alpha^{new}$ obtained by a straightforward calculation:
\begin{equation}
\alpha^{new}=\alpha+\frac{1}{4}-\frac{1}{4}\left(  \left(  3-2i-j\right)
^{2}-4(i-2)(i+j-1)\right)  \text{, }\delta=\frac{1-2i-j}{2}\text{ .}%
\end{equation}
For the critical value obtained for $\alpha^{new}=1/4$, one obtains
$\Psi(a)=a^{\frac{3-2i-j}{2}}\ln(a/a_{0})$ where $\frac{3-2i-j}{2}$ is always
positive in the chosen ranges for $i$ and $j$.\ Again, $\Psi^{new}(a)$ is
real, then also $\Psi(a)$ is real.

\bigskip

\textit{A numerical example:} For larger values of $a$, the terms proportional
to $\Lambda$ and $k$ become important. It is then instructive to study a
numerical case which is reminiscent of the potentials studies in\ Refs.
\cite{hh,hawking2,vilenkin,gibbons}, with the inclusion of the\ additional
short-range $-1/a^{2}$ potential. To this end, we start by rewriting the WDW
equation in natural units and in terms of the dimensionless $a^{\prime
}=a/\sqrt{G}$:%

\begin{equation}
-\frac{\mathrm{d^{2}}\Psi}{\mathrm{d}a^{\prime2}}+\text{ }\tilde{V}%
_{eff}(a^{\prime})\Psi=0\text{ where }\tilde{V}_{eff}(a^{\prime})=\tilde
{k}a^{\prime2}-\lambda a^{\prime4}-\frac{\alpha}{a^{\prime2}}\text{,}
\label{wdwad}%
\end{equation}
with dimensionless constants $\tilde{k}=\frac{9\pi^{2}k}{4}$, $\lambda
=\frac{3\pi^{2}}{4}\Lambda G,$ and $\alpha$ already introduced in Eq.
(\ref{alfa}). In general, we could not find analytic solutions of this
equation and we did not derive the conditions of self-adjointness of this new
operator. However, the behavior of the wave function close to $a^{\prime}=0$
should be anyway dominated by the $1/a^{\prime2}$ term and thus we will again
have solutions which vanish in $a^{\prime}=0$ and are real, see Fig. 1 for an
illustrative example.


It is easy to prove that the wave function is $L^{2}(0,+\infty)$: indeed in
the limit $a^{\prime}\rightarrow\infty$, Eq. (\ref{wdwad}) admits solution of
the type $\sqrt{a^{\prime}}\,\mathrm{BesselJ}(\pm1/6,\sqrt{\lambda}a^{\prime
3}/3)$ which scales as $1/a^{\prime}$ at infinity. We thus find a solution
which interpolates between the analytical one in Eq. (\ref{psisol}) and the
one of Hartle and Hawking (real and normalizable).

\begin{figure}[ptb]
\begin{centering}
		\epsfig{file=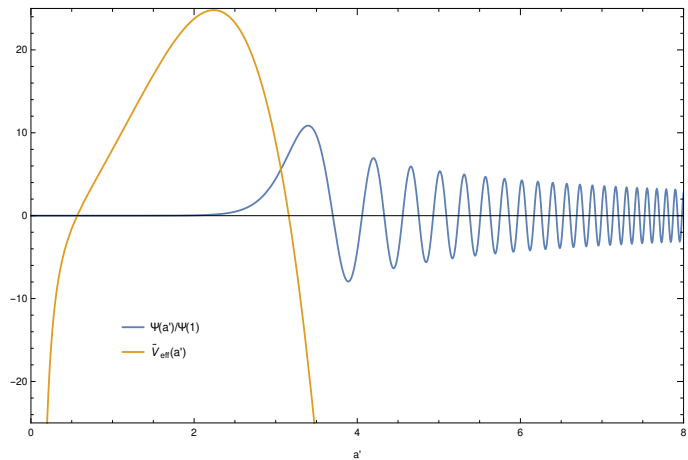,height=6cm,width=9cm,angle=0}
		\caption{Example of one specific choice of the parameters of the effective potential $\tilde{V}_{eff}$ ($\tilde k=10$, $\lambda=1$, $\alpha=1$) and the corresponding wave function solution of the WDW equation (normalized to the value at $a'=1$.) }
	\end{centering}
\end{figure}

\textit{Conclusions: }In this work, we have studied the effect of a stiff
matter component in the very early phase of the Universe. The corresponding
potential in the WDW equation is proportional to $-1/a^{2}.\ $This very
interesting and atypical potential has some remarkable features for the WDW
equation: the wave function vanishes at the origin, is defined only for
positive $a$, and is real. Moreover, our qualitative results do not depend on
a large class of choices of the operator ordering of the WDW equation. In the
future, more detailed and more realistic numerical studies which take into
account such an interaction dominated gas as well as additional terms are
needed. The investigation of possible phenomenological implication of an
initial stiff matter on the early inflation and on present cosmological
observables represent a promising outlook of the present work.

\end{document}